# Free-running SIMilarity-Based Angiography (SIMBA) for simplified anatomical MR imaging of the heart

John Heerfordt[1,2], Kevin K. Whitehead[3], Jessica A.M. Bastiaansen[1], Lorenzo Di Sopra[1], Christopher W. Roy[1], Jérôme Yerly[1,4], Bastien Milani[1], Mark A. Fogel[3], Matthias Stuber[1,4], Davide Piccini[1,2]

[1]Department of Diagnostic and Interventional Radiology, Lausanne University Hospital and University of Lausanne, Lausanne, Switzerland

[2]Advanced Clinical Imaging Technology, Siemens Healthcare AG, Lausanne, Switzerland

[3]Division of Cardiology, Department of Pediatrics, The Children's Hospital of Philadelphia, University of Pennsylvania Perelman School of Medicine, Philadelphia, USA

[4]Center for Biomedical Imaging (CIBM), Lausanne, Switzerland

**Submitted to Magnetic Resonance in Medicine**

**Grant support:**

M.S.: Schweizerischer Nationalfonds zur Förderung der Wissenschaftlichen Forschung, Grant/Award Number: Project grant #173129

J.A.M.B.: Schweizerischer Nationalfonds zur Förderung der Wissenschaftlichen Forschung, grant number PZ00P3_167871. Swiss Heart foundation, grant number FF18054. Emma Muschamp foundation.

**Corresponding author:**

Davide Piccini PhD, Center for Biomedical Imaging, Centre Hospitalier Universitaire Vaudois, Rue du Bugnon 46, BH 7.82, 1011 Lausanne, Switzerland. E-mail: piccinidavide@gmail.com, Twitter: @CVMR_Lausanne

# Abstract

**Purpose**

Whole-heart MRA techniques typically target pre-determined motion states and address cardiac and respiratory dynamics independently. We propose a novel fast reconstruction algorithm, applicable to ungated free-running sequences, that leverages inherent similarities in the acquired data to avoid such physiological constraints.

**Theory and Methods**

The proposed SIMilarity-Based Angiography (*SIMBA*) method clusters the continuously acquired k-space data in order to find a motion-consistent subset that can be reconstructed into a motion-suppressed whole-heart MRA. Free-running 3D radial datasets from six ferumoxytol-enhanced scans of pediatric cardiac patients and twelve non-contrast scans of healthy volunteers were reconstructed with a non-motion-suppressed regridding of all the acquired data (*All Data*), our proposed *SIMBA* method, and a previously published free-running framework (*FRF*) that uses cardiac and respiratory self-gating and compressed sensing. Images were compared for blood-myocardium interface sharpness, contrast ratio, and visibility of coronary artery ostia.

**Results**

Both the fast *SIMBA* reconstruction (~20s) and the *FRF* provided significantly higher blood-myocardium sharpness than *All Data* (*P*<0.001). No significant difference was observed among the former two. Significantly higher blood-myocardium contrast ratio was obtained with *SIMBA* compared to *All Data* and *FRF (P<0.01)*. More coronary ostia could be visualized with both *SIMBA* and *FRF* than with All Data (*All Data*: 4/36, *SIMBA*: 30/36, *FRF*: 33/36, both *P*<0.001) but no significant difference was found between the first two.

**Conclusion**

The combination of free-running sequences and the fast *SIMBA* reconstruction, which operates without *a priori* assumptions related to physiological motion, forms a simple workflow for obtaining whole-heart MRA with sharp anatomical structures.

# INTRODUCTION

High-resolution MRI of the whole heart plays an important role in the assessment of congenital heart disease (1), including malformations at the coronary origins (2) and anatomical measurements of the aorta and great vessels (3). It also allows for the examination of coronary artery stenoses (4,5) and for planning of subsequent scans using highly detailed 3D images (6). To obtain motion-suppressed volumetric anatomical images, a specific motion state of the heart – e.g. the diastolic cardiac resting phase at end-expiration – is generally targeted during the data acquisition. In state-of-the-art approaches, cardiac and respiratory motion components are commonly regarded as separate entities for which different motion suppression strategies are applied. Artifacts originating from *cardiac motion* can be minimized by triggering the acquisition to specific cardiac phases using an external ECG signal. The drawbacks of this approach include: the placement of the electrodes, the requirement for careful planning of appropriate trigger delays and acquisition windows, the fact that ECG signals are unreliable in the MRI environment, and that the RR interval may vary during the acquisition (7). In turn, *respiratory motion artifacts* can be minimized using breath holds or by gating the data acquisition to pre-defined respiratory levels with respiratory bellows (8,9) or with diaphragmatic navigator echoes (10). However, while the latter also requires dedicated planning and is both time-inefficient and sensitive to variations in the breathing pattern (11), the former implies a compromise in resolution and spatial coverage, especially in uncooperative patients.

Whole-heart MRA approaches that address these shortcomings have been proposed. Examples include: avoiding ECGs by utilizing cardiac self-gating (12,13), avoiding diaphragmatic navigators by performing respiratory motion-correction based on 1D projections (14) or image-navigators (15), and using physiological self-gating signals to bin the data into different motion states in order to resolve motion (12,13,16,17). Free-running whole-heart pulse sequences (12,18) that acquire data continuously, irrespective of the ongoing physiological motion, allow for simplified planning and become particularly powerful when combined with binning since cine images can be obtained (12,13,17,18). Recently, a fully automated free-running framework (*FRF*), comprising continuous acquisition, full self-gating, and cardiac and respiratory motion-resolved reconstruction, was proposed (13).

All these methodologies, hinge on targeting specific motion states by extracting and processing cardiac and respiratory motion independently, either prospectively during the acquisition or retrospectively as part of the reconstruction routine. This builds on the assumption that data from such targeted motion states must intrinsically have a high degree of motion-consistency, thus resulting in images with a low level of motion artifacts. Nonetheless, there remain a whole range of challenges. For instance, when a pre-determined motion state is targeted, the diastolic resting phase at end-expiration is often assumed to be optimal. This may not always be the case as in certain subjects end-systole (19) and other respiratory levels than end-expiration (20) may be more adequate. An additional complicating factor is that the diastolic cardiac resting phase may be affected by temporal changes in the heart rate during a scan, which ideally should be accounted for by the imaging technique (21). Cardiac and respiratory self-gating methods typically rely on explicit assumptions regarding expected frequency ranges for cardiac and respiratory motion (12,13). Lastly, motion-resolved compressed sensing reconstructions may last several hours (17). With all this in mind, what if we reverse the underlying assumptions so that, instead of regarding cardiac and respiratory motion components as separate entities and targeting specific motion states, we try to directly leverage intrinsic similarities in the acquired data to independently find a subset with a high level of motion consistency that can quickly be reconstructed into a motion-suppressed image?

To that end, a novel fast similarity-based whole-heart image reconstruction method is proposed, described, validated, and tested in this work. On the acquisition side, existing free-running data collection strategies (12,13,18,22) that offer uniform coverage of k-space over time (via reordering schemes based on the golden angle (23,24)) are used. On the reconstruction side, the acquired MR imaging data are clustered based on their inherent similarities in order to identify motion-consistent subsets that can be used for reconstructing motion-suppressed static whole-heart MRA. Consequently, there is no need to explicitly differentiate motion sources from one another nor use iterative reconstruction provided that the subsets are populated sufficiently. By combining existing free-running acquisitions with the proposed reconstruction method, we hypothesize that: 1) whole-heart images with sharp anatomical image features, including coronary arteries as a surrogate endpoint for motion consistency, can be reconstructed using similarity-based clustering free of *a priori* physiological assumptions; and that 2) the selected data intrinsically originate from well-defined phases of the cardiac and respiratory cycles. Preliminary results from this work have been presented in abstract form (25).

# THEORY

## Background

Consider the task of reconstructing a sharp static 3D anatomical image of the heart using MR data obtained with an ungated and untriggered free-running acquisition, where the scan time was long enough for considerable cardiac and respiratory motion to occur. Without any motion suppression strategy, it is clearly sub-optimal to reconstruct all the acquired data into one single image, as this would be heavily degraded by motion artifacts (26). Therefore, we seek to identify a subset of the collected data that was acquired during a similar, unspecified anatomical motion state and that can consequently be used for reconstructing an image with minimal motion artifact degradation.

## Similarity-based clustering

The method proposed in this work, SIMilarity-Based Angiography (*SIMBA*), involves grouping the acquired MR data into $k$ disjoint subsets/clusters $\mathbf{C} = \{C_1, C_2,\ldots, C_k\}$ that are populated based on the similarity of a total number of $T$ reference data vectors $\mathbf{s}_1, \mathbf{s}_2, \ldots, \mathbf{s}_T, \mathbf{s}_i \in \mathbb{R}^N$. These vectors are constructed from certain MR imaging data, intrinsically modulated by physiological motion, that are acquired throughout the whole free-running acquisition at regularly spaced time points $t_1, t_2,\ldots, t_T$ (cf. Figure 1A). Candidates for such reference data can be found in the self-gating literature, e.g. the magnitude of the k-space center coefficient (27), 1D projection images in the superior-inferior (SI) direction (14) or a vectorized representation of an image-navigator (28). Hence, if the reference data vectors from some of the time points show a high degree of similarity, it may be inferred that the heart may have been in a similar position and contractile state at those instances. Note that $t_{i+1} - t_i$ should be short enough that it can be assumed that very little organ motion occurred during that period.

For the concrete assessment of similarity, let us create a matrix $\mathbf{S} \in \mathbb{R}^{N \times T}$ by concatenating the $T$ reference data vectors $\mathbf{s}_1, \mathbf{s}_2, \ldots, \mathbf{s}_T$ with $N$ elements each according to their temporal order of acquisition. Thus, each column of $\mathbf{S}$ corresponds to a reference data vector. If a free-running sequence repeatedly acquires the same k-space line or sample at the time points $t_1, t_2,\ldots, t_T$, such a matrix can easily be created by letting these measurements serve as reference data. Once $\mathbf{S}$ has been populated, the task at hand is to group its columns with reference data vectors (i.e. points in an $N$-dimensional space) into the different clusters $C_1, C_2,\ldots, C_k$ using an appropriate clustering technique and a distance metric. If the reference data vectors are of high dimensionality, dimensionality reduction before the clustering step may be beneficial (29). In this work we employ k-means (30), meaning that the data points are assigned to clusters in a way that minimizes the sum of the intra-cluster sum-of-squares distances. Formally, this can be expressed as $\arg\min_{\mathbf{C}} \sum_{j=1}^{k} \sum_{\mathbf{s_i} \in C_j} \|\mathbf{s}_i - \boldsymbol{\mu}_j\|^2$ where $\mathbf{s}_i$ is the i$^{th}$ column of **S** and $\boldsymbol{\mu}_j$ is the centroid of the j$^{th}$ cluster which corresponds to the average of all the data points assigned to the cluster. This optimization problem is normally solved in an iterative fashion sensitive to initial conditions, which is why ascertaining a good spread of the initial centroids with the *k-means++* algorithm (31) may be advantageous.

### From clusters to images

After assigning the reference data obtained at the time points $t_1$, $t_2$,…, $t_T$ to different clusters, the remaining k-space data have to be assigned as well. For interleaved acquisitions where one reference data vector is obtained per interleaf, this is easily accomplished by assigning all the data within one interleaf to the same cluster as its reference data. Alternatively, one can imagine assigning all the data acquired within a certain temporal window around each time point to the same cluster as the corresponding reference data vector. Provided that the different clusters contain motion-consistent data with a relatively uniform k-space distribution, standard image reconstruction of the data in such a cluster should result in an image with a low level of both motion- and sampling-related artifacts.

# METHODS

## SIMBA for free-running 3D radial whole-heart imaging

Here, we describe and validate an implementation of *SIMBA* designed for ungated free-running sequences with interleaved 3D radial sampling patterns (12,18). Such acquisitions are suitable for *SIMBA* because they both make use of a golden-angle increment assuring that a random subset of interleaves provides approximately uniform k-space sampling and acquire an SI-readout every interleaf. Here, we use 1D Fourier transforms of such SI-readouts for creating the reference data vectors that form the data matrix **S** in the *SIMBA* clustering (One SI-readout per interleaf is acquired). Given a free-running acquisition of the heart consisting of a total number of $N_{interleaves} \in \mathbb{N}$ radial interleaves, our goal is to group these interleaves into *k* disjoint clusters $C_1, C_2,\ldots, C_k$, $k << N_{interleaves}$ using reference data vectors created from the SI-projections from the time points $t_1, t_2, \ldots, t_{N_{interleaves}}$. Each reference data vector, is a concatenation of 1D gradients (with $N_{samples}$ elements per gradient) of the corresponding SI-projections from $N_{coil} = 4$ fixed and centrally located chest surface coil elements. Gradients of the projections, computed as first order forward differences in the readout direction, are used to reduce the dependency on the sequence contrast due to the intrinsic normalization and to highlight interfaces such as the lung-liver interface. The main steps of this *SIMBA* implementation are the following (Figure 1):

1. Data acquisition with an interleaved free-running 3D radial sequence with regularly acquired SI-readouts.
2. 1D gradients of the magnitude SI-projections from the selected chest coil elements are computed. Each gradient vector is standardized to have a mean value of zero a variance of one before concatenating the gradients into reference data vectors for the different interleaves. Subsequently, these reference data vectors are organized into the data matrix $\mathbf{S} \in \mathbb{R}^{(N_{coil} \times N_{samples}) \times N_{interleaves}}$.
3. Dimensionality reduction with principal component analysis (PCA) is used to reduce the dimensionality of **S**. The $N_{PC} = 20$ principal components that explain the most variance are extracted by applying PCA along the interleaf direction, computed with a singular value decomposition $\mathbf{S} = \mathbf{U\Sigma V}^T$ where $\mathbf{U} \in \mathbb{R}^{(N_{coil} \times N_{samples}) \times N_{interleaves}}$, $\mathbf{\Sigma} \in \mathbb{R}^{(N_{coil} \times N_{samples}) \times N_{interleaves}}$, $\mathbf{V} \in \mathbb{R}^{N_{interleaves} \times N_{interleaves}}$. The choice of $N_{PC} = 20$ was empirically made as a trade-off between explanatory power and restricting the dimensionality. The first principal component is discarded and the remaining $N_{PC-1} = 19$ principal components are used to create the low-dimensional data matrix $\mathbf{\hat{S}} \in \mathbb{R}^{N_{PC-1} \times N_{interleaves}}$.

4. Given this low-dimensional representation $\hat{s}_1, \hat{s}_2, ..., \hat{s}_{N_{\text{interleaves}}}, \hat{s} \in \mathbb{R}^{N_{\text{PC}}-1}$ of the different interleaves ($\hat{s}_j$ denotes the j$^{\text{th}}$ column of $\hat{S}$), k-means clustering is used to group the interleaves into the *k* disjoint clusters $C_1, C_2, ..., C_k$. The number of clusters is determined based on existing knowledge from triggered 3D radial coronary MRA, namely that 12,000-15,000 readouts are adequate for reconstructing a whole-heart volume with isotropic millimetric resolution (32,33) without the need for an expensive iterative reconstruction approach. Seeking to maintain approximately that amount of data in the most populated cluster $C_{\text{largest}}$, an automated search procedure was developed, which picks *k* in the range 10-14 based on which particular *k* that minimizes the average distance between the centroid and the data points of the most populated cluster, i.e. $\underset{k \in \{10,11,...,14\}}{\arg\min} = \frac{1}{|C_{\text{largest}}|} \sum_{\hat{s} \in C_{\text{largest}}} \|\mu_{\text{largest}, k} - \hat{s}\|_2$. Here $\mu_{\text{largest}, k}$ denotes the centroid of the most populated cluster after performing *k*-means clustering with *k* clusters and its cardinality $|C_{\text{largest}}|$ the number of interleaves in that cluster.

5. Finally, the data corresponding to the interleaves of the most populated cluster $C_{\text{largest}}$ are extracted and a standard non-Cartesian reconstruction is performed, consisting of: density compensation, non-uniform fast Fourier transform (NUFFT) (34) (regridding to a Cartesian grid, inverse Fourier transformation, and roll-off correction (35)), and sum-of-squares coil combination.

## Data acquisition

To evaluate the performance of the *SIMBA* algorithm for different contrasts and populations, a heterogeneous set of data was used in this IRB approved *in vivo* study (study populations and acquisition parameters are found in Table 1). In total, 18 free-running datasets were acquired using three different types of free-running sequences (12,18,22), all using the same prototype 3D radial phyllotaxis trajectory (36). Each acquisition consisted of 5181-5749 radial interleaves with 22-24 readouts each. Six pediatric cardiac patients (6 ± 4 years, 4 male) underwent imaging with contrast-enhanced gradient echo (CE-GRE) under respiratory ventilation after administration of ferumoxytol (37), 6 healthy volunteers (27 ± 3 years, 5 male) with fat-saturated bSSFP (FS-bSSFP) (18), and another 6 healthy volunteers (27 ± 2 years, 4 male) with a fast-interrupted steady state (FISS) (22) protocol. Data were acquired on a 1.5T MAGNETOM Avanto$^{\text{fit}}$ (6xCE-GRE) and a 1.5T MAGNETOM Aera (6xFS-bSSFP, 4xFISS) and a 3T MAGNETOM Prisma$^{\text{fit}}$ (2xFISS) clinical scanners (Siemens Healthcare, Erlangen, Germany). The ECG was recorded as a reference for retrospective analysis of *SIMBA*'s data selection, but was neither used during image acquisition nor reconstruction.

## Image reconstruction

Each of the free-running datasets was reconstructed with *SIMBA* and two reference techniques: a conventional 3D gridding reconstruction (34) of all the collected radial data without motion correction (*All Data*) and a fully self-gated cardiac and respiratory motion-resolved compressed sensing reconstruction that recently was reported as part of *FRF* (13). The motion-resolved *FRF* reconstructions used four respiratory phases and a temporal cardiac resolution of 50 ms. For all comparisons, end-expiratory *FRF* images from the diastolic resting phase were used. In addition, the reconstruction times were measured. The non-iterative *All Data* and *SIMBA* reconstructions can be performed on a standard computer whereas powerful hardware is required for the iterative motion-resolved compressed sensing reconstruction algorithm *XD-GRASP* (17,38) used in *FRF*.

## Analysis of image quality

To test our first hypothesis, image quality was evaluated using different metrics. The sharpness of the blood-myocardium interface was measured for all three reconstruction types. This was done by fitting parametrized sigmoid functions to the blood-myocardium interface of the left ventricle, using a parametrization where the value of the slope parameter describing the sigmoid depends on the interface's sharpness (39). These measurements were performed in coronal reformats at the same anatomical position for all the reconstruction types. The final sharpness value for each image type and subject was obtained as the average slope parameter of six fitted sigmoid functions, three oriented in the medial-lateral and three in the SI-direction. The blood-myocardium contrast ratio (CR) was quantified by comparing average signal intensities in regions of interest (one in the left ventricle's blood pool and one in the septal wall of the myocardium) in a mid-myocardial axial reformat. The blood-myocardium CR was computed as $\text{CR} = \frac{\text{Signal}_{blood} - \text{Signal}_{myocardium}}{\text{Signal}_{myocardium}}$. Because of the intrinsic contrast differences among the three sequences, the CR for *All Data* and *SIMBA* were statistically compared as a ratio to the corresponding values from *FRF*. For *All Data* and *SIMBA*, also measurements of blood signal-to-noise ratio (SNR) and blood-myocardium contrast-to-noise ratio (CNR) were performed. To correct for the noise distribution in magnitude images, the normalization factors in (40) were used. CNR and SNR were not measured in *FRF* images since both parallel imaging and compressed sensing impact the background noise's distribution.

Furthermore, the visibility of the ostia of the right coronary artery (RCA) and the left main (LM) were categorized as either visible or non-visible in a consensus reading by two authors (JH and DP with 3 and 10 years of experience in coronary MRA, respectively). Subsequently, the visible length and the sharpness (in proximal 4 cm and full visible course) of both the right coronary artery (RCA) and the combined LM + left anterior descending (LAD) coronary arteries were quantified using the Soap-Bubble software (41). Since the visibility of the coronary arteries typically is very low in motion-degraded reconstructions from *All Data*, this analysis was only performed for *SIMBA* and *FRF* in vessels where both methods could visualize the ostia.

### Analysis of the *SIMBA* data selection

To test our second hypothesis, we characterized the automatically selected data in S*IMBA*. The number of selected interleaves was ascertained and the ratio with respect to the total number of interleaves was computed for each dataset. Moreover the uniformity of the 3D radial sampling in k-space was analyzed by examining the distribution of the selected readouts on the unit sphere. Specifically, the average great-circle distance between readouts and their four closest neighbors as well as the relative standard deviation (RSD) thereof were computed. This RSD metric reflects how variable the distances between the k-space samples and their neighbors are, with a lower value corresponding to higher uniformity (36). These two analyses were also performed for *All Data* and *FRF*.

We further examined the provenance of the data selected by *SIMBA* relative to their cardiac phases and respiratory positions. In the cardiac dimension, the selected readouts were categorized as either systolic or diastolic based on their associated ECG-timestamps, corresponding to how long after an R-wave each readout was acquired. As a reference, individual systolic durations were obtained by approximating the QT interval using an empirical formula (21,42), and then subtracting an approximate QR duration of 50 ms (43). Based on this categorization, the average percentages of data that originated from the subject-specific systolic and diastolic intervals were computed. Additionally, the *SIMBA* data selection as a whole was categorized as systolic, diastolic or mixed (<75% of the data originated from the most populated phase) for each subject. In the respiratory dimension, we investigated in which respiratory state the readouts selected by *SIMBA* were acquired. For that analysis, the respiratory self-gating signal from *FRF* (13) served as a subject-specific reference to sort all the acquired imaging data into four different respiratory levels. For every subject, the percentage of data selected by *SIMBA* that was contained in the two most end-expiratory bins and whether the majority of the data originated from expiration or inspiration were determined.

### Statistical analysis

For statistical comparisons of continuous variables, either two-sided paired sample t-tests or Wilcoxon signed-rank tests were used depending on whether the variable appeared to be normally distributed or not. Jarque-Bera tests (44) were used for assessing normality. McNemar's tests were used for nominal variables. In all statistical testing, $P < 0.05$ was being considered statistically significant. To correct for multiple comparisons, Bonferroni correction was used where applicable.

## Testimonials from clinicians

As part of this translational research, we provided this technology to clinicians at one of our institutions. A free-running CE-GRE protocol using ferumoxytol and inline *SIMBA* reconstruction was implemented on a clinical MRI system (<6 minutes acquisition time and ~1 minute reconstruction time). Motion-resolved *FRF* images were reconstructed retrospectively when requested. After initial evaluation in congenital heart disease patients (not part of this study), including both children and adults, we solicited testimonials from three clinicians using the open-ended question:

"*What does SIMBA mean for your clinical work, what are the advantages and drawbacks and do you have requirements and/or wishes for its future development?*"

# RESULTS

### Image reconstruction

*SIMBA* allowed for fast 20-second reconstructions of 3D whole-heart images: PCA: 0.5 ± 0.1 seconds, selection of the number of clusters: 9.0 ± 3 seconds, final k-means clustering 2.0 ± 1.4 seconds, and NUFFT and sum-of-squares coil combination 7.4 ± 1.4 seconds. On the same computer, 35.3 ± 6.0 seconds were required only for the NUFFT and coil combination of *All Data*. The iterative *FRF* reconstructions took 3:18 ± 1:13 hours per subject, although it should be emphasized that those reconstructions generated 5D images with 32-88 (heartrate dependent) different 3D volumes and were performed on a more powerful machine.

### Image quality

Visually, *SIMBA* suppressed most of the motion blur seen in the *All Data*-reconstructions and provided comparable image quality to *FRF* (Figure 2). Both *SIMBA* and *FRF* provided significantly higher average blood-myocardium sharpness than *All Data* (both $P < 0.001$), but no significant difference was observed between the two ($P = 0.31$ before Bonferroni correction, Figure 3A). The average sharpness values (i.e. sigmoid slope parameter) were: *All Data* 1.6 ± 0.2, *SIMBA* 2.5 ± 0.4, *FRF* 2.4 ± 0.5. The blood-myocardium CR was 1.0 ± 0.8 for All Data, 1.5 ± 1.2 for *SIMBA* and 1.0 ± 0.7 for *FRF* (Figure 3B). *SIMBA* provided significantly higher CR than both *All Data* and *FRF* (both $P < 0.01$ after Bonferroni correction, Figure 3C). *All Data* images had a higher blood SNR and blood-myocardium CNR than *SIMBA* (SNR: 20.1 ± 9.1 and 11.4 ± 5.5 and CNR: 9.4 ± 7.1 and 6.5 ± 5.3, respectively, both $P < 0.001$).

Significantly more coronary ostia were visible with both *SIMBA* and *FRF* compared to *All Data* (both $P < 0.001$, *All Data*: 4/36, *SIMBA*: 30/36, *FRF*: 33/36). No significant difference was found between *SIMBA and FRF* ($P=0.25$ before Bonferroni correction). The average vessel sharpness of the RCA was significantly lower with *SIMBA* compared to *FRF* but no significant difference was observed for LM+LAD (Table 2). On average, a significantly longer part of the LM+LAD was seen with *SIMBA* than with *FRF* ($P < 0.05$) while no significant difference was observed for the RCA. Examples of coronary artery reformats are provided in Figure 4.

### SIMBA data selection

The cluster selected by *SIMBA* ($C_{largest}$) contained on average 14,354 ± 2,238 readouts which corresponded to 11.4 ± 1.7% of the acquired data and 25.6 ± 4.3% of the radial Nyquist limit (45). Images of one subject and the corresponding selected and rejected reference data vectors are depicted in Figure 5. The average great-circle distance was 7.4·$10^{-3}$ ± 2.3·$10^{-5}$ for *All Data*, 1.8·$10^{-2}$ ± 1.2·$10^{-3}$ for *SIMBA* and 4.9·$10^{-2}$ ± 6.9·$10^{-3}$ for *FRF*. This average distance was significantly smaller for *SIMBA* when compared to *FRF* ($P < 0.001$), but significantly higher when compared to *All Data* ($P < 0.001$). On average, the data selected using *SIMBA* had a more uniform distribution in k-space than the more undersampled *FRF* frames, this according to the measured RSDs of 27.6 ± 3.4% and 30.5 ± 2.6%, respectively ($P < 0.01$). The oversampled *All Data* reconstructions (~215% Nyquist) showed the highest uniformity with an RSD of 5.9 ± 0.01% (both $P < 0.001$).

Regarding the physiological motion consistency of the *SIMBA* selection, on average 69 ± 22% of the selected data originated from diastole. Diastolic images were obtained in 8/18 subjects, systolic in 1/18, and a combination thereof in 9/18 (Figure 6). This combination typically contained early systolic and diastolic data and was not associated with poor image quality (the CE-GRE examples in Figure 2 and Figure 5 are such cases). In the subject with the highest average heartrate in this study, an infant with 146 beats per minute, considerable cardiac motion blur was seen although the data was only selected from a limited portion of the cardiac cycle (Figure 7A). Regarding breathing motion, the selected cluster contained mainly end-expiratory data. On average 88.3 ± 22.9% of the selected data were found in the two most end-expiratory *FRF* bins. In one subject, *SIMBA* selected inspiratory data and a motion-degraded reconstruction was obtained (Figure 7B). Figure 8 shows images from one pediatric patient reconstructed from the data in different clusters. Interestingly, the clusters appear to correspond to different motion states, with e.g. the most populated cluster corresponding to a dilated left ventricle and the second most populated to a contracted one.

### Testimonials from Clinicians

All three clinicians provided testimonials that are listed in supporting document S1.

# DISCUSSION

This paper aimed at the development and validation of the *SIMBA* image reconstruction which, when combined with various free-running sequences, provides a simplified solution for anatomical whole-heart MRI. The novelty of the proposed similarity-based method consists of: automatically targeting an arbitrary subject-specific motion state, not making assumptions about the regularity and periodicity of the ongoing physiological processes, and not requiring the isolation of different motion sources from one another.

The findings that *SIMBA*-images showed higher blood-myocardium sharpness and contrast ratio than *All Data*-images and similar blood-myocardium and LM+LAD sharpness as *FRF*-images, suggest that motion-suppressed whole-heart MRA can be obtained without making *a priori* physiological assumptions. Non-significant differences in sharpness between *SIMBA* and *FRF* might be explained by the fact that one *SIMBA* image, reconstructed without compressed sensing, contains more data than one *FRF* frame (~3% Nyquist), but *FRF*'s regularization means that information is shared among frames. That may also explain that a longer part of the LM+LAD could be seen with *SIMBA*. The finding that the fast-moving RCA was sharper in *FRF* images might be attributed to the temporal cardiac bin width of 50 ms, which appears smaller than what was obtained with *SIMBA* (cf. Figure 6). The higher SNR and CNR in reconstructions with *All Data* compared to *SIMBA* might be attributed to the considerably lower background noise in the *All Data* images, a consequence of the 9-fold amount of data. Additionally, *SIMBA* improved the visualization of the coronary ostia compared to *All Data*, which can be explained by more motion-consistent data. This is corroborated by the physiological analyses which showed that mainly end-expiratory data from limited portions of the cardiac cycle were selected. Interestingly, non-evident ways of combining data from non-adjacent phases in the cardiac cycle were identified by *SIMBA* as seen when early systolic and mid-diastolic data were clustered together (Figure 6). This result is intuitively explained by the heart having a similar anatomical shape during those two phases of the cardiac cycle. The blurriness of the heart in some of the *SIMBA* reconstructions (e.g. the infant in Figure 7A and the FISS example in Figure 2) is likely a result of residual cardiac motion in the selected data. One possible explanation is that data were clustered on an interleaf-by-interleaf basis. This may sometimes result in too long continuous portions of data being used for reconstruction of the *SIMBA* image. To mitigate this issue, fewer readouts per interleaf could be acquired or the number of clusters increased to improve the motion-consistency, at the expense of increased undersampling.

Although the acquisition time is similar to that of conventional ECG-triggered and navigator-gated whole-heart MRA sequences and the output remains a 3D image, there are important advantages of a free-running sequence with *SIMBA* reconstruction. These include that there is no need to place ECG leads, to carefully setup acquisition windows, nor to plan respiratory navigators. *SIMBA* might also be less sensitive to RR-variability and respiratory drift since no specific motion state is targeted, though this remains to be investigated. An important future study will be to compare the image quality obtained with *SIMBA* to that of conventional methods. Concerning the reconstruction times, the computationally inexpensive components of the *SIMBA* algorithm allowed for fast image reconstruction. Hence one use case is immediate display of *SIMBA* images at the scanner console combined with a more comprehensive motion-resolved *FRF* dataset being made available to the reader as soon as the longer reconstruction process has finished.

## Study limitations

The datasets used in this study were acquired with three different free-running sequences. However, since the reconstruction methods were applied to the very same rawdata, we do not consider this to limit the value of the reported results. It is also desirable that a fixed *SIMBA* implementation can handle heterogeneous data. Moreover, although the *SIMBA* concept does not require specifying expected physiological frequencies, the periodicity of the cardiac and the respiratory motions ensures that there are enough motion-consistent data for performing reconstructions without motion correction. While motion-consistent static images can be reconstructed with *SIMBA*, there is by design no one-to-one correspondence with the underlying physiology. Therefore, the extraction of pre-specified systolic or diastolic images is not straightforward at this juncture. A limitation in the data analysis is that the coronary artery sharpness of *SIMBA* and *FRF* was compared only in the vessels where both methods could visualize the ostia, which does not take into account that *FRF* could visualize more coronary ostia overall.

## Future directions

Optimization of the type of reference data vectors, dimensionality reduction method and clustering technique would most likely improve the resulting image quality. So would integrating techniques such as coil sensitivity weighted coil-combination, parallel imaging, compressed sensing and/or denoising. One could imagine using information from all the readouts in an interleaf for creating a reference data vector, e.g. by reconstructing a sub-image (28). Alternatively, the sampling rate of the reference data vectors could be increased by using center-of-k-space coefficients or an external Pilot Tone (46). Moreover, *k*-means enforces spherical clusters which might be sub-optimal. Further improvement may be achieved by weighting the data points within the selected cluster based on their similarity or applying intra-cluster motion correction. Furthermore, the proposed implementation only used the data in the most populated cluster for image reconstruction. As illustrated in Figure 8, different clusters correspond to different motion states. Hence, exploiting spatial correlations with a motion-resolved *XD-GRASP* (38) approach might enhance the images and potentially provide dynamic information. Identifying clusters corresponding to maximal left ventricular contraction and dilation might allow for obtaining left ventricular ejection fraction. *SIMBA* might also be useful in other applications, e.g. for identifying respiratory motion-consistent data in late gadolinium enhancement imaging (47) or in imaging of the abdomen or the lungs as well as for reconstructing an anatomical reference image in 5D flow imaging (48,49).

# CONCLUSION

We successfully tested our first hypothesis, that whole-heart MRA images with sharp anatomical features including visible coronary arteries can be reconstructed from different types of free-running acquisitions using the data-driven *SIMBA* technique. Therefore, it can be inferred that *SIMBA* effectively minimizes respiratory and cardiac motion without requiring specific *a priori* assumptions related to physiology. Moreover, we demonstrated that the similarity-based clustering algorithm intrinsically tends to select data that originate from limited portions of the cardiac cycle at end-expiration, thus corroborating our second hypothesis. The *SIMBA* technique allows for simplified whole-heart imaging because it is computationally inexpensive, time efficient, and is applicable to free-running acquisitions that do not require any gating, triggering or complicated planning.

## A) Overview of the SIMilarity-Based Angiography (SIMBA) concept

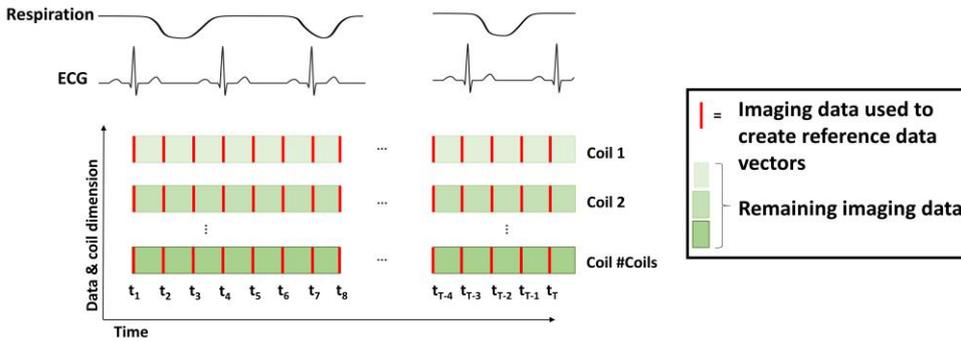
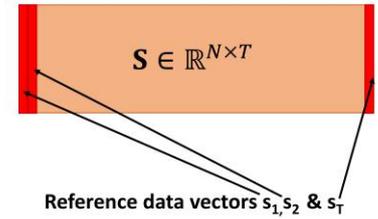
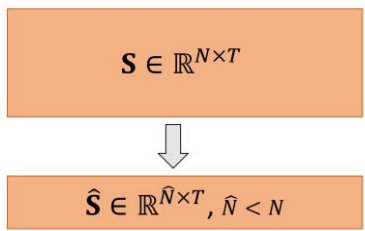
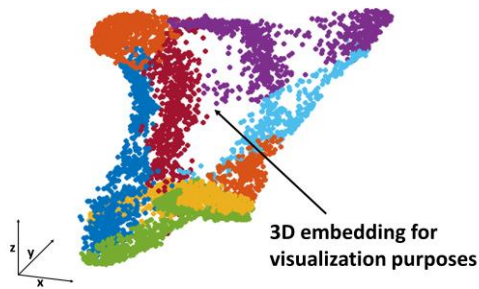
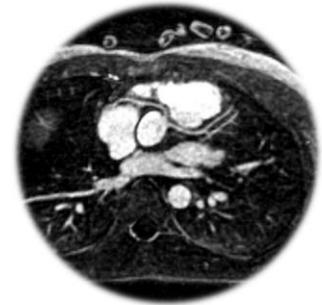

## B) Choices for implementation tested in this work

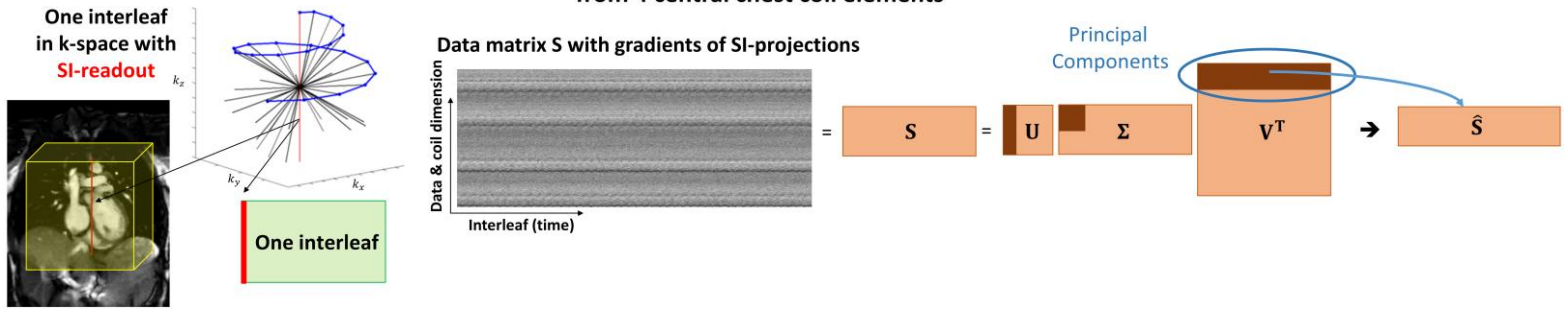

*Figure 1. SIMilarity Based Angiography (SIMBA)*

*A) From the imaging data of a free-running acquisition, a number of reference data vectors are first constructed and then fed to a clustering algorithm. By design, the reference data vectors are modulated by physiological motion which means that a cluster of similar reference data vectors is expected to correspond to data that were acquired during a similar anatomical state.*

*B) The specific implementation tested in vivo in this work is designed for interleaved free-running 3D radial sequences with regularly acquired readouts in the superior-inferior (SI) direction. Computing the Fourier transform of such readouts yields projection images whose gradients are used to create reference data vectors. Prior to the actual clustering, PCA is used to reduce the dimensionality.*

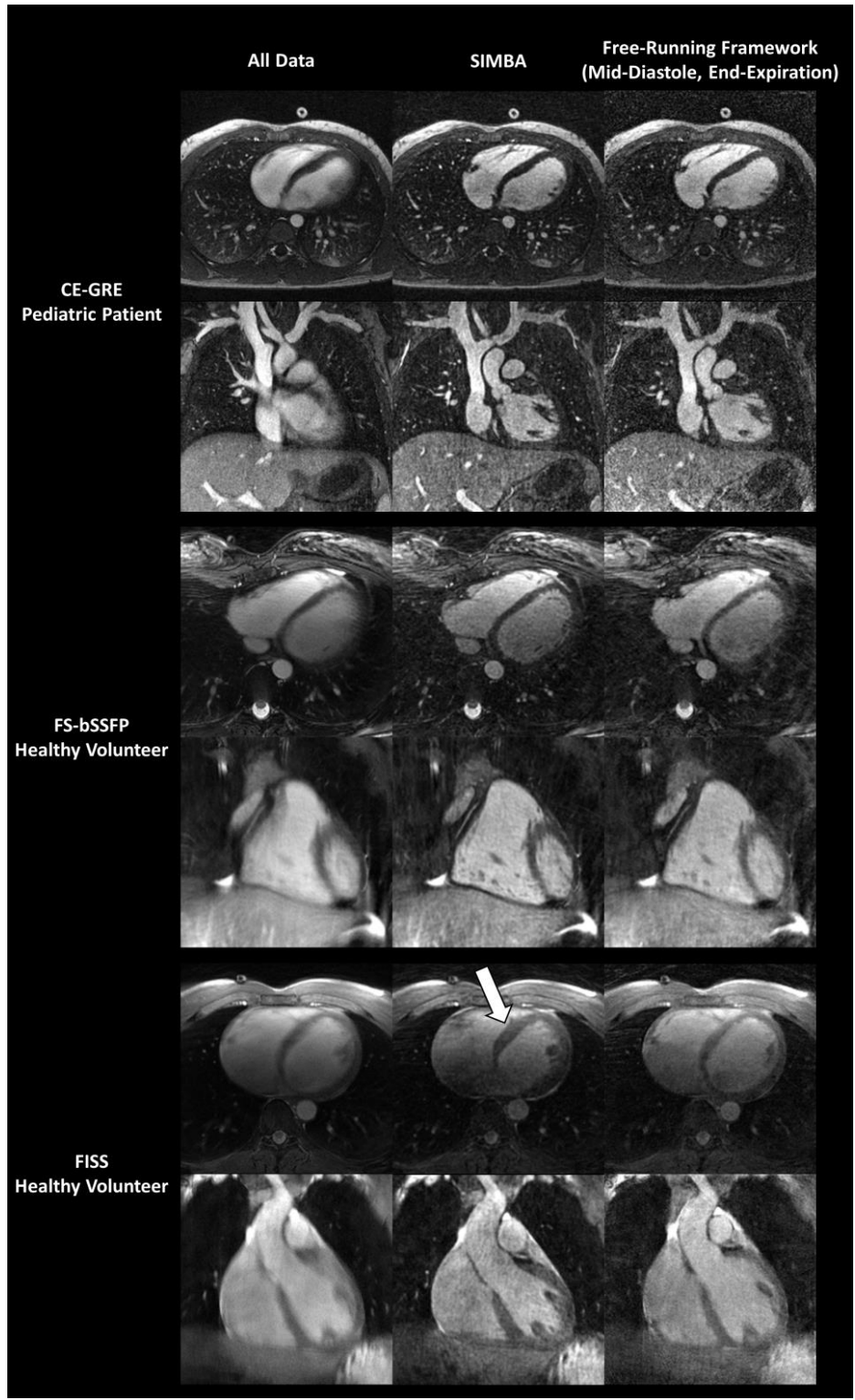

*Figure 2. Representative example images, at locations where the coronary arteries generally are visible, from three different subjects scanned with different types of free-running 3D radial acquisitions and reconstructed with: All Data (left column), SIMBA (middle column) and FRF (right column). In general, the SIMBA images appear sharper than their All Data counterparts. In the CE-GRE and FS-bSSFP examples, the motion-state selected by SIMBA matches that of the FRF images quite well. However, in the FISS example, SIMBA extracted data corresponding to a more contracted left ventricle (arrow).*

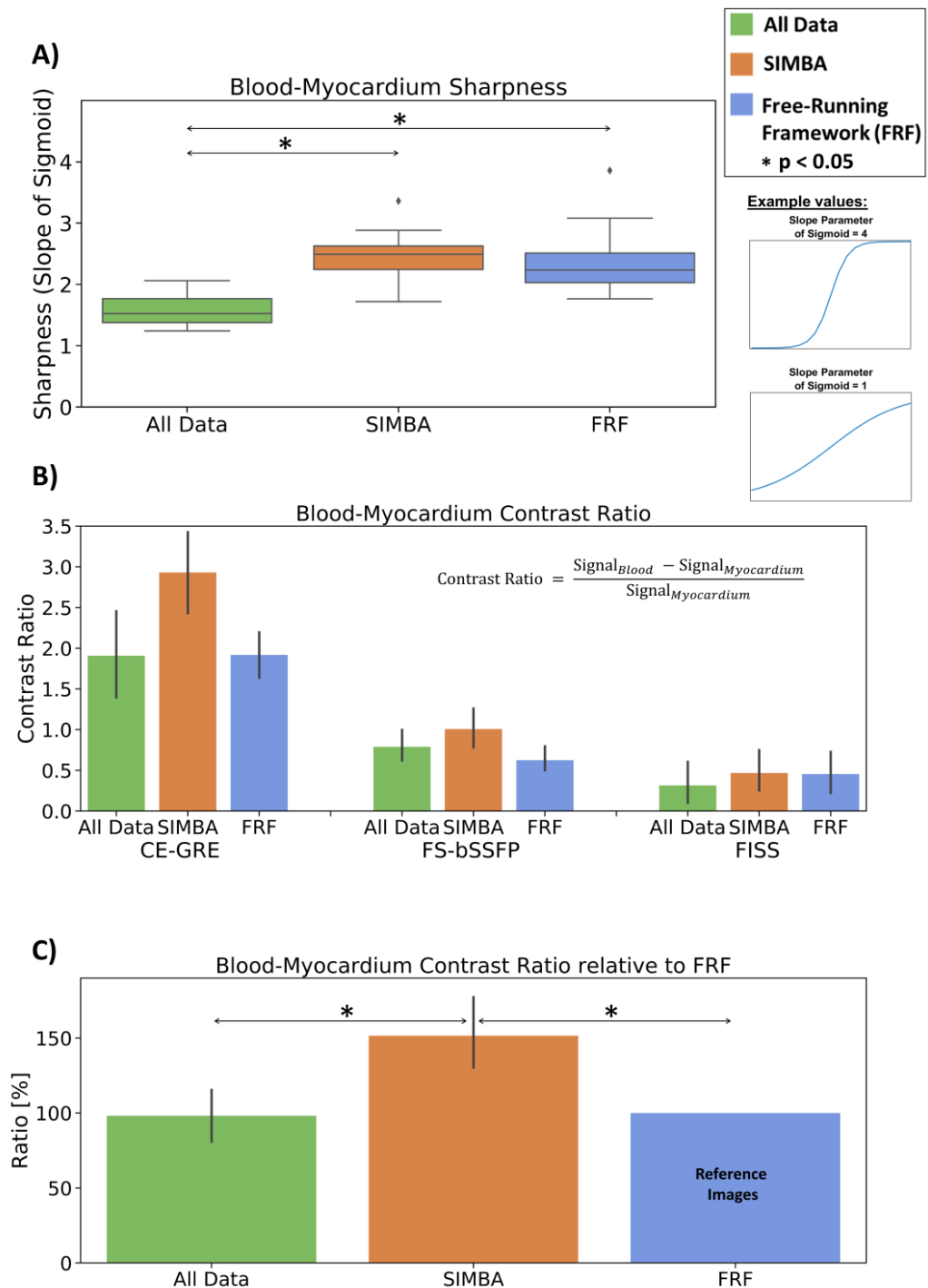

*Figure 3. Blood-Myocardium Analyses*

**A)** *The blood-myocardium sharpness obtained from fitting parametrized sigmoid functions to the interface and using the average slope parameter value as a measure of sharpness. As expected, SIMBA and FRF both provided better sharpness than the motion-corrupted All Data reconstructions.*

**B)** *Bar plots showing the average blood-myocardium contrast ratio with confidence intervals. As expected, CE-GRE provided higher contrast than the non-contrast sequences.*

**C)** *To allow for statistical comparisons with N=18 and aggregating contrast-enhanced and non-contrast data, normalization was performed by expressing the individual subjects' contrast ratio values as a percentage of the corresponding FRF values. SIMBA provided significantly better blood-myocardium contrast than the two other approaches.*

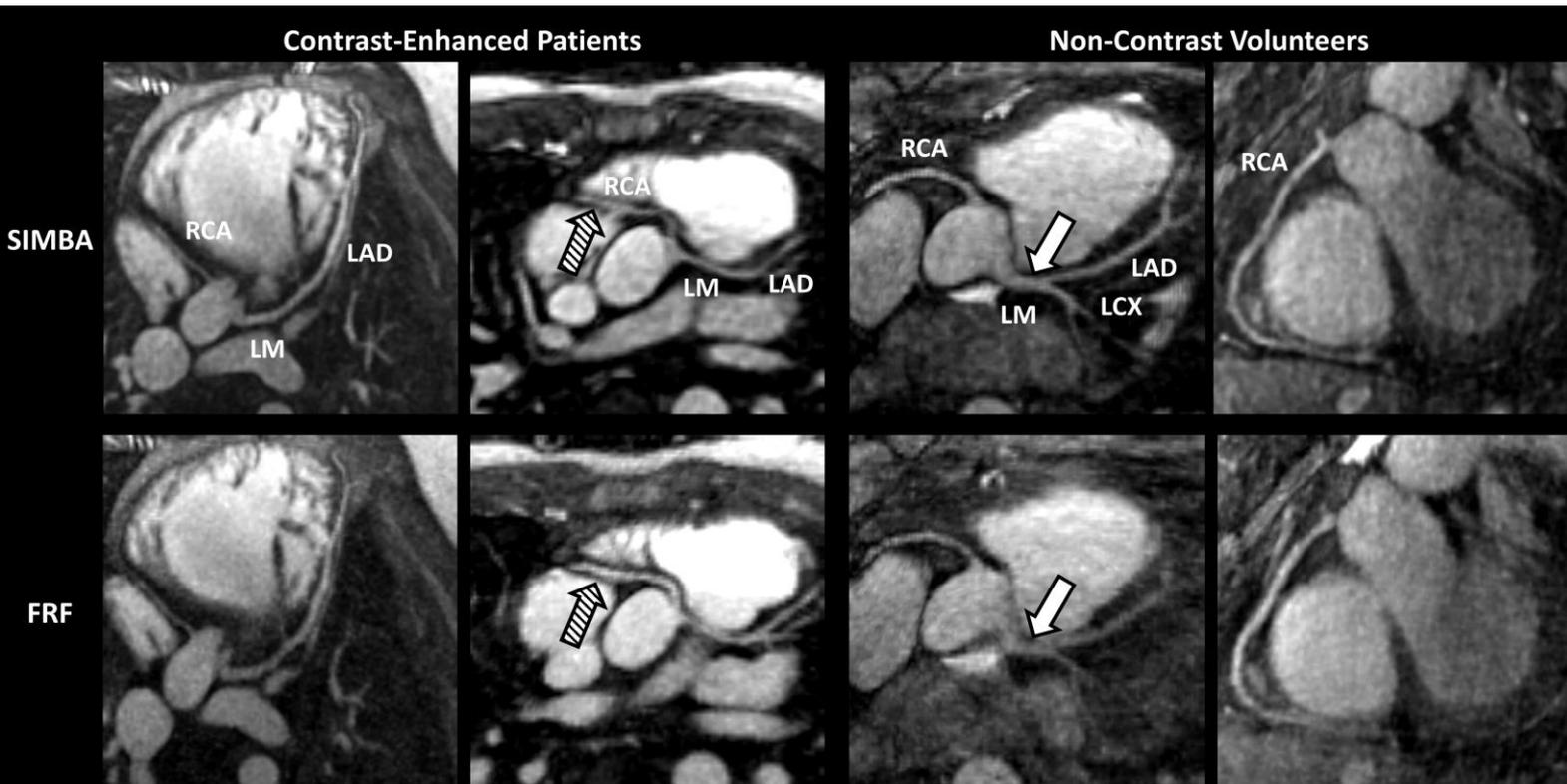

*Figure 4. Examples of multiplanar reformats depicting coronary arteries from four different subjects. SIMBA and FRF images appeared relatively similar overall (cf. leftmost patient and rightmost volunteer). However, in the pediatric patient with an anomalous ostium of the RCA, where both the left main and the RCA originate from the left coronary cusp, FRF provided better right coronary artery vessel conspicuity (striped arrows). Conversely, in the leftmost one of the two volunteers, SIMBA visualized the left coronary tree more clearly (solid arrows).*

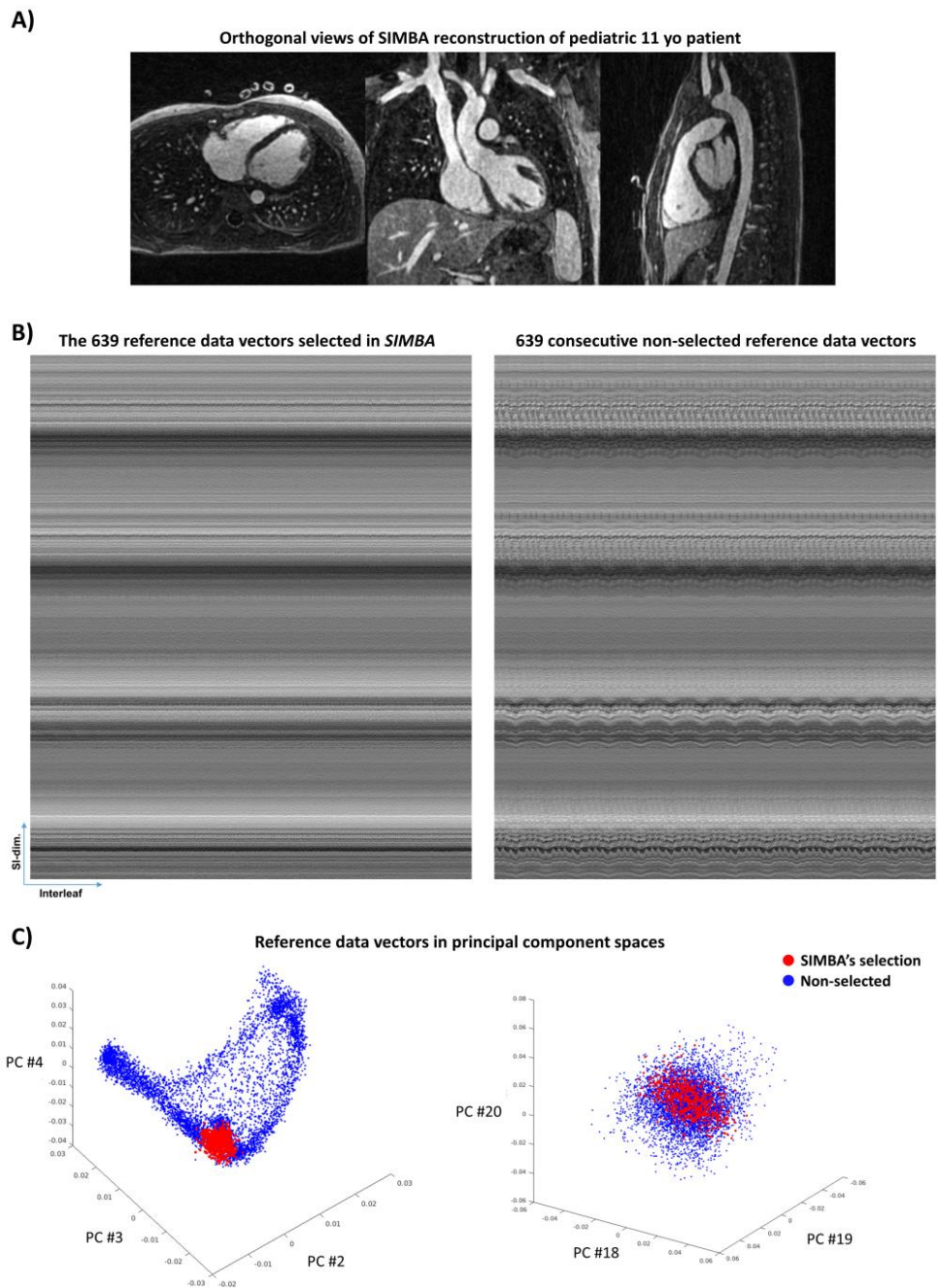

*Figure 5. Example of a SIMBA reconstruction and the corresponding data selection*

*A) SIMBA reconstruction of a CE-GRE scan of a pediatric patient. The visibility of the coronary arteries and the sharp lung-liver interface point towards a selection of motion-consistent data, in agreement with subfigure B.*

*B) Stacked gradients of the SI-projections in the selected cluster (left) and an equal amount of consecutive non-selected interleaves (right). The reference data vectors in the selected cluster show less evidence of physiological motion.*

*C) When plotting the reference data vectors of the selected interleaves in 3D spaces of different principal components, a general observation included a very clear discrimination between the selected (red) and the non-selected (blue) data in lower principal component dimensions (left) whereas in the higher dimensions, it was more mixed (right).*

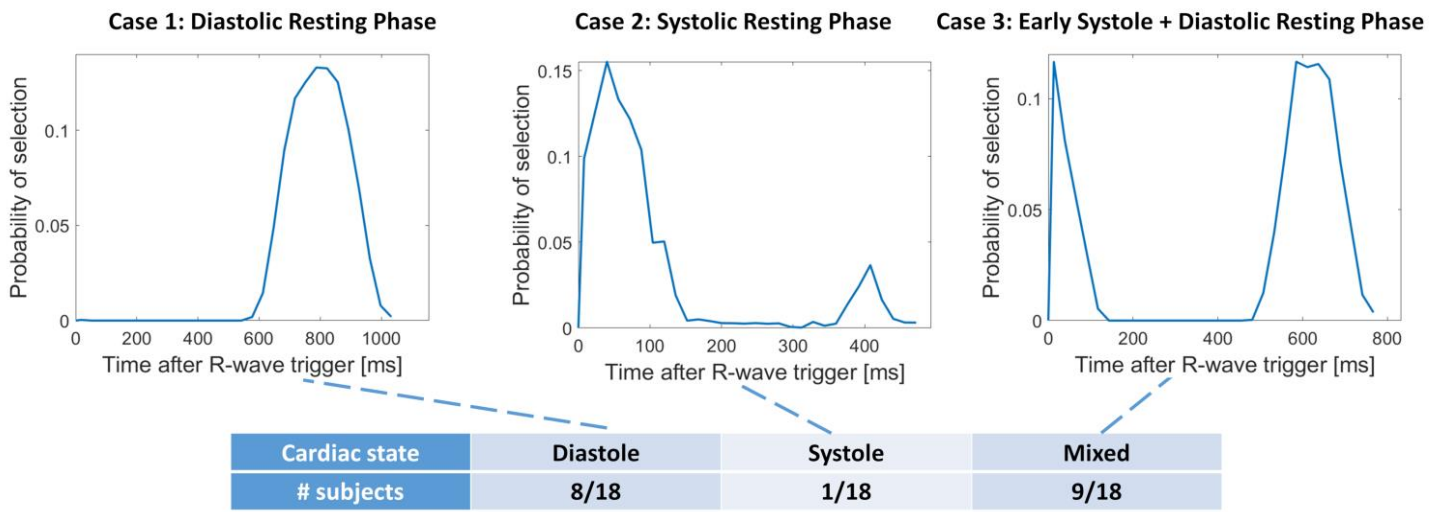

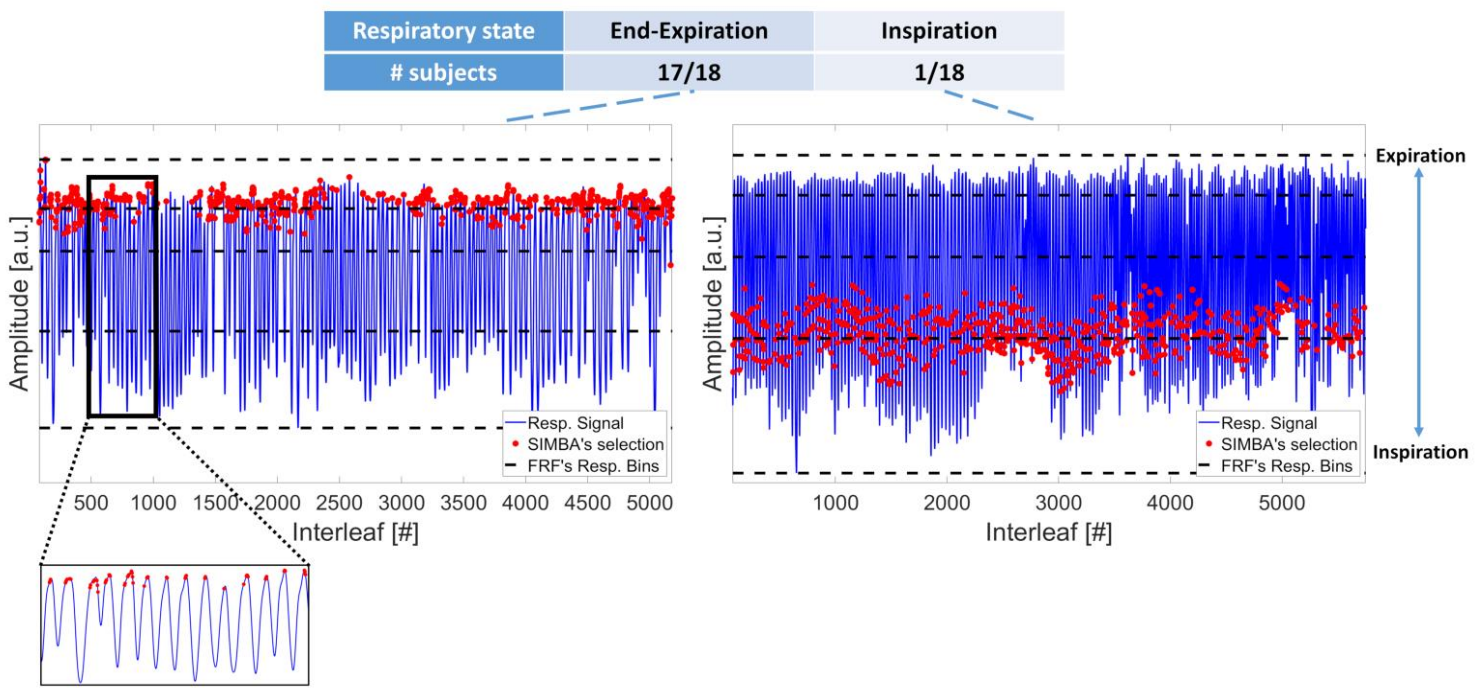

*Figure 6. Physiological analysis of SIMBA's data selection*

*A) Three main cardiac data selection scenarios occurred with SIMBA: diastolic data (Case 1), systolic data (Case 2) or a combination of early systolic and diastolic data (Case 3). The latter is interesting since the heart is likely to have a similar shape during those two time points in the cardiac cycle, but it does not correspond to how one would intuitively extract data from an ECG.*

*B) SIMBA mainly selected data from the most common respiratory state, end-expiration. The zoom depicts the respiratory self-gating signal and overlaid selection in more detail. In one subject, SIMBA selected inspiratory data and in that case the image quality was poor, consistent with the wider spread of the data points.*

SIMilarity-Based Angiography (SIMBA)            Heerfordt et al.

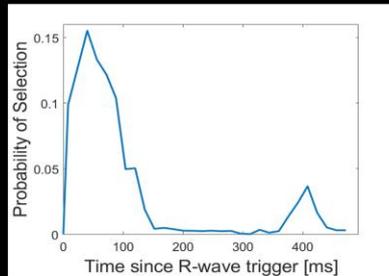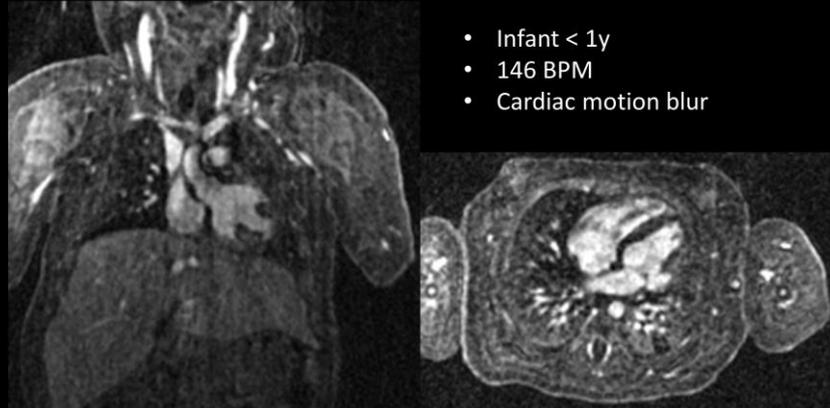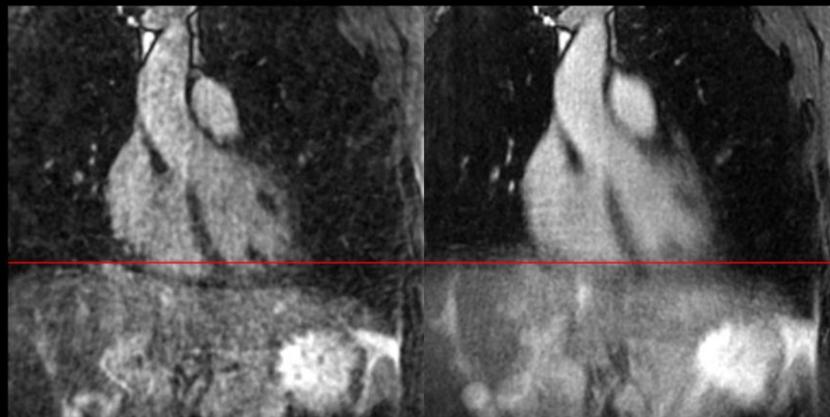

*Figure 7. Example of two subjects in whom SIMBA did not perform satisfactorily.*

*Example A: In this infant with an average heartrate of 146 beats per minute (BPM), the SIMBA reconstruction showed considerable cardiac motion blurring.*

*Example B: In this healthy volunteer, SIMBA selected inspiratory data (cf. liver dome with respect to the red reference line, same subject as bottom right respiratory signal in Figure 6B) which probably contributed to the overall blurriness.*

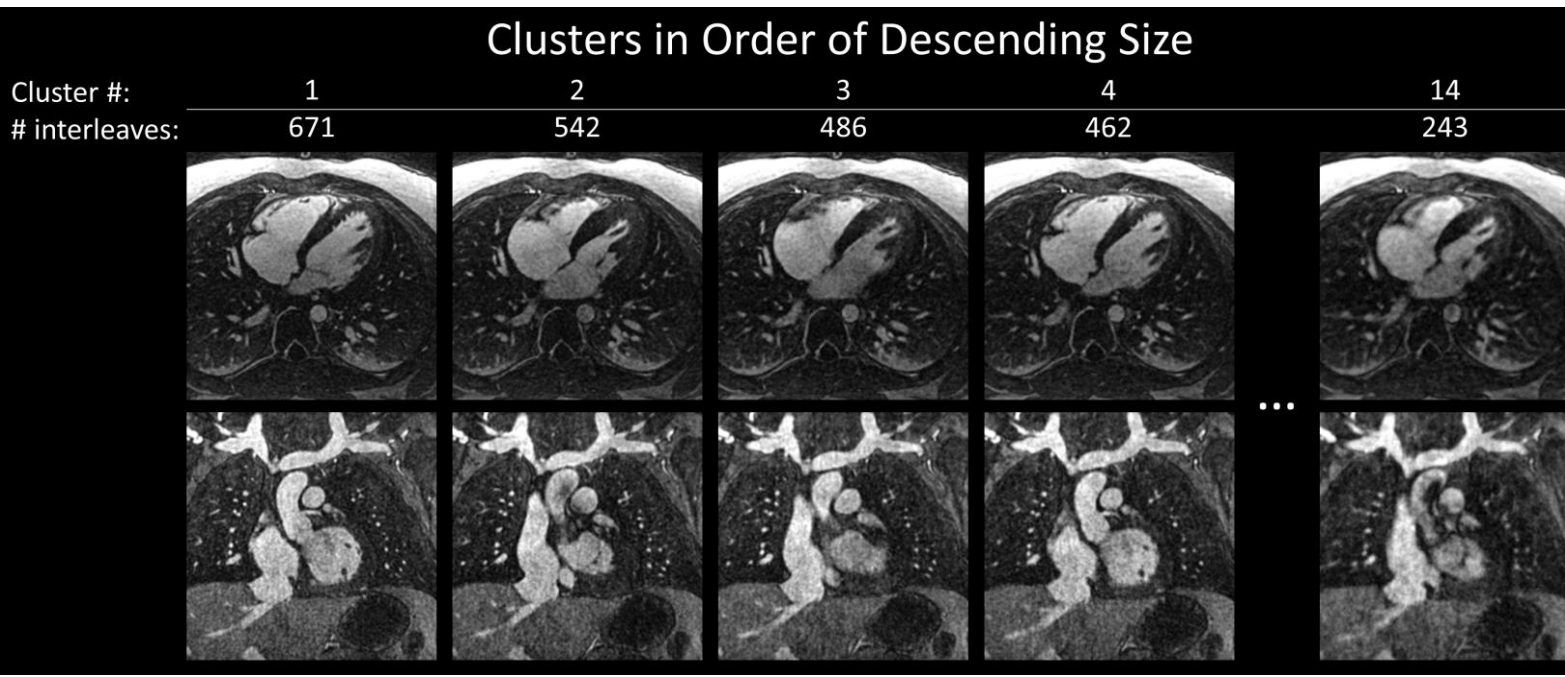

*Figure 8. Throughout this paper only the most populated of the different clusters identified for each subject has been used for image reconstruction. However, as exemplified here in a 9 year old male patient with ferumoxytol, reconstructions of additional, less populated clusters may also provide informative images of high quality. Interestingly, in this case, it appears like the largest and the second largest cluster correspond to diastolic and systolic images, respectively.*

*Table 1. Datasets used in study*

| Parameter | CE-GRE Patient Scans | FS-bSSFP Volunteer Scans | FISS Volunteer Scans |
|---|---|---|---|
| Cohort | 6 pediatric cardiac patients | 6 healthy adult volunteers | 6 healthy adult volunteers |
| Scanner | 1.5T MAGNETOM Avanto[fit] | 1.5T MAGNETOM Aera | 4 on 1.5T MAGNETOM Aera, 2 on 3T MAGNETOM Prisma[fit] |
| Sequence | GRE | Fat-saturated bSSFP with 10 ramp-up pulses | Fast-interrupted steady state (FISS) |
| Contrast agent? | Ferumoxytol (4 mg/kg) | Non-contrast | |
| Sampling scheme | 3D radial Spiral Phyllotaxis | | |
| Number of spokes | 126,478 (5749 interleaves with 22 spokes) | | 124,344 (5181 interleaves with 24 spokes) |
| Field-of-view [mm$^3$] | $180^3$-$220^3$ | $220^3$ | $220^3$ |
| Acquired resolution [mm$^3$] | $0.9^3$-$1.1^3$ | $1.1^3$ | $1.1^3$ - $1.5^3$ |
| Echo time/Repetition time [ms] | 1.55-1.74/ 3.34-3.65 | 1.56/3.12 | 1.3-1.5/2.6-3.0 (1.5T), 1.7/3.4 (3T) |
| Radio frequency excitation angle | 15º | 90º | 40º-60º |
| Acquisition time [minutes:seconds] | 7:03 -7:42 | 14:17 | 6:55-10:37 |

*Table 2. Assessment of the conspicuity of the coronary arteries*

| Data<br>Metric | All Data | SIMBA | Free-Running Framework (diastolic resting phase, end expiration) |
|---|---|---|---|
| Visible LM ostium [count] | 2/18 | **15/18** | **15/18** |
| Visible RCA ostium [count] | 2/18 | 15/18 | **18/18** |
| Analyses performed in the vessels with ostia visible with both *SIMBA and FRF:* | | | |
| LM+LAD Visibility [cm] | N/A | **8.3 ± 3.5†** | 6.8 ± 2.4 |
| Proximal LM+LAD Sharpness [%] | N/A | **40.3 ± 9.2** | 38.1 ± 9.1 |
| Full LM+LAD Sharpness [%] | N/A | **37.2 ± 7.9** | 35.9 ± 7.3 |
| RCA Visibility [cm] | N/A | 6.8 ± 3.2 | **7.8 ± 3.0** |
| Proximal RCA Sharpness [%] | N/A | 34.8 ± 9.7 | **39.4 ± 9.6†** |
| Full RCA Sharpness [%] | N/A | 34.2 ± 9.2 | **39.2 ± 7.1†** |

**bold text** = highest mean/count

† = $P < 0.05$

## Abbreviations

BPM = Beats Per Minute

CE-GRE = Contrast-Enhanced Gradient Echo

CNR = Contrast-to-Noise Ratio

CR = Contrast Ratio

FISS = Fast-Interrupted Steady State

FRF = Free-Running Framework

FS-bSSFP = Fat-Saturated balanced Steady-State Free Precession

LAD = Left Anterior Descending Coronary Artery

LCX = Left Circumflex coronary artery

LM = Left Main coronary artery

NUFFT = Non-Uniform Fast Fourier Transform

SI = Superior-Inferior

SIMBA = Similarity-Based Angiography

SNR = Signal-to-Noise Ratio

RCA = Right Coronary Artery

RSD = Relative Standard Deviation

## Acknowledgements

The authors would like to thank Dr Estelle Tenisch, Dr Tobias Rutz, and Dr Milan Prsa at Lausanne University Hospital for providing testimonials on their initial experience in patients using ferumoxytol-enhanced free-running imaging with SIMBA reconstruction implemented on a clinical MRI system.

# Supporting Information S1 – Testimonials from Clinicians

In answer to the question:

"What does *SIMBA* mean for your clinical work, what are the advantages and drawbacks and do you have requirements and/or wishes for its future development?"

the clinicians provided the following answers:

**Pediatric radiologist:**

"Concerning pediatric patients, this technique is particularly interesting because it has a very good spatial resolution even in small hearts.

The shortness of the acquisition time is a huge advantage for young patients that we do not sedate because we can start the exam with *SIMBA* and we know that we will obtain all relevant information even if the child does not want to stay in the machine longer than 15 minutes. Even in sedated patients, who can wake up at any moment, the *SIMBA* gives us the security of having a complete exam, avoiding rescheduling failed exams due to failed sedation. *SIMBA* works without apnea, which is very useful in sedated children who breathe freely.

Moreover, we think that those exams done with the *SIMBA* will permit to lower the need for sedation or general anesthesia, saving stress for the parents and the child as well as saving machine usage time and consequently money.

For me, as a pediatric radiologist, it is very thrilling to see those images of the heart because I foresee all the potential utilization of the sequence in the rest of the body."

**Cardiologist 1:**

"The first experiences with the new *SIMBA* technique are extremely fascinating. This technique allows to easily acquire a 3D whole heart providing the whole anatomical information. I run it at the beginning of the scan for the planning of the further sequences like e.g. flow sequences by providing at the same time high resolution anatomical information. The high quality of the sequence already allowed to identify a postoperative complication in a patient which had not been identified on previous MRI exams and which will actually have an impact on the patient's treatment.

The possibility to obtain also function makes the sequence extremely interesting and attractive as in only a couple of minutes we get both, anatomy and cardiac function which usually takes at least 20 minutes. This will allow to shorten scan time by rendering the scan much more comfortable for the patients as we are independent of the ECG signal and breath-holds.

These promising results encourage continuing with the project and I am very honored and happy to participate in the development of the sequence.

I am looking forward to implement the sequence into my routine scan protocol. It would be great to have reconstructions of specific phases of the cardiac cycle and of the whole 3D cine directly on the scanner."

**Cardiologist 2:**

"The *SIMBA* technique with ferumoxytol has the potential to transform how we do cardiac MRI in congenital heart disease.

Currently, a standard exam lasts 1 hour, requires general anesthesia in small children, and high-resolution 3D imaging of the heart and thoracic vessels is not consistently feasible.

With *SIMBA* + ferumoxytol, we can consistently acquire high-resolution 3D images of the heart and thoracic vessels with superior image quality in just 6 minutes. This reduces scan time significantly, eliminates the need for general anesthesia, and obviates the need for irradiating CT scans. The images are of such high quality to allow for a very detailed and precise evaluation of the most complex pre- and post-operative anatomy of congenital heart disease. Moreover, cardiac and respiratory motion-resolved images can be obtained from the same acquisition, further improving diagnostic accuracy. With the potential of adding measurements of ventricular volumes, this sequence is likely to have a profound impact on clinical decision-making and patient management.

What we need to further improve the utility of the sequence is faster reconstruction of the cardiac and respiratory motion-resolved images, and reduction of noise in the images."